**Hallucinations in Organization-backed AI advisors: Evidence about Skepticism, Verification, and Reliance in Goal-Directed Use**

Simon J. Blanchard[a,*], Aaron M. Garvey[b], Laura O'Laughlin[c]

[a] McDonough School of Business, Georgetown University, 37th and O Streets NW, Washington, DC 20057, USA
[b] Haslam College of Business, University of Tennessee, 916 Volunteer Boulevard, Knoxville, TN 37996, USA
[c] Analysis Group, Montreal, Quebec, Canada

** Corresponding author. Email: sjb247@georgetown.edu*

## Abstract

Generative artificial intelligence (GenAI) systems are increasingly used by organizations to deliver information to consumers, patients, students, employees, and citizens. These systems can hallucinate, producing plausible but inaccurate responses. A central question for AI-advised decisions is therefore not only whether users rely on inaccurate information, but whether they recognize that a response may require verification. To answer this question, we review emerging empirical evidence relevant to hallucination detection in goal-directed interactions, with a focus on organization-backed AI advisors. We distinguish three constructs that existing studies often conflate: whether users are skeptical of information presented, whether they verify it (distinguishing attempted from successful verification), and whether the result of verification affects reliance on the information. Across studies examining product search, medical decision-making, content generation, and chatbot-assisted tasks, several patterns emerge. Nearly all studies measure reliance, while variables such as user skepticism and verification of the information are more often targeted by an intervention than measured directly. The cues used to prompt scrutiny of the AI response are predominantly related to the AI output, such as source citations. Whereas general warnings about the risk of hallucinations are among the most deployable interventions for organizations, they show the weakest and most mixed effects in the studies reviewed. More targeted cues about where AI systems tend to fail appear promising, however evidence remains limited. Although the existing literature posits that users may be more likely to scrutinize responses related to particular areas of content, no studies varied the content category, leaving this question open for further research. In future research, measuring skepticism and verification separately from reliance may clarify what current evidence shows, what it only implies, and which questions require further exploration.

## 1. Introduction

Organizations increasingly use generative artificial intelligence (GenAI) systems as tools to assist customers, patients, students, employees, and citizens in completing ordinary tasks such as obtaining information about a reservation. Despite their utility, such systems can produce plausible but inaccurate responses. When incorrect information is provided by an AI tool, do users notice and act on the problem while pursuing their goals?

Our review examines the existing literature on interactions between individuals and a conversational AI that involve a risk of hallucination. We use the natural-language-generation term "*hallucinated information*" to refer to plausible generated output that is unfaithful to the relevant source, policy, evidence base, or decision context [1]. We define "*organization-backed AI advisors*" as systems offered through an organizational channel or in a context in which users reasonably treat the system as providing information on the organization's behalf (e.g., service chatbots, medical assistants, employer knowledge tools). In these settings, users are typically engaged in task completion, such as obtaining health advice or finding a consumer product, rather than explicit evaluation of the system's accuracy.

For organization-backed AI advisors, hallucination detection can be usefully separated into three constructs: skepticism, verification, and reliance. We define skepticism as recognizing that a response may be inaccurate or require verification [2]; verification as the act of checking the accuracy of the response against one or more other sources [3], a construct with two layers because verification may be attempted without successfully detecting an error; and reliance as whether the response affects the user's decision [4,5]. These constructs are distinct: users may recognize that an answer requires checking yet still rely on it, or reject an answer without verifying a specific error. The distinction follows work in trust in automation and appropriate reliance, which separates attitudes toward systems from behavioral use [4,5,6].

Skepticism may arise through at least two pathways. Output-based skepticism is prompted by features of the response itself, such as implausibility, inconsistency, or conflict with prior knowledge. Category-based skepticism is prompted by knowledge that certain content types, such as numerical claims, policy provisions, medical guidance, and citations, might be more error-prone or higher-stakes. Once skepticism arises, the user may attempt to verify the answer; verification may or may not succeed, and its result may or may not change reliance (see Figure 1).

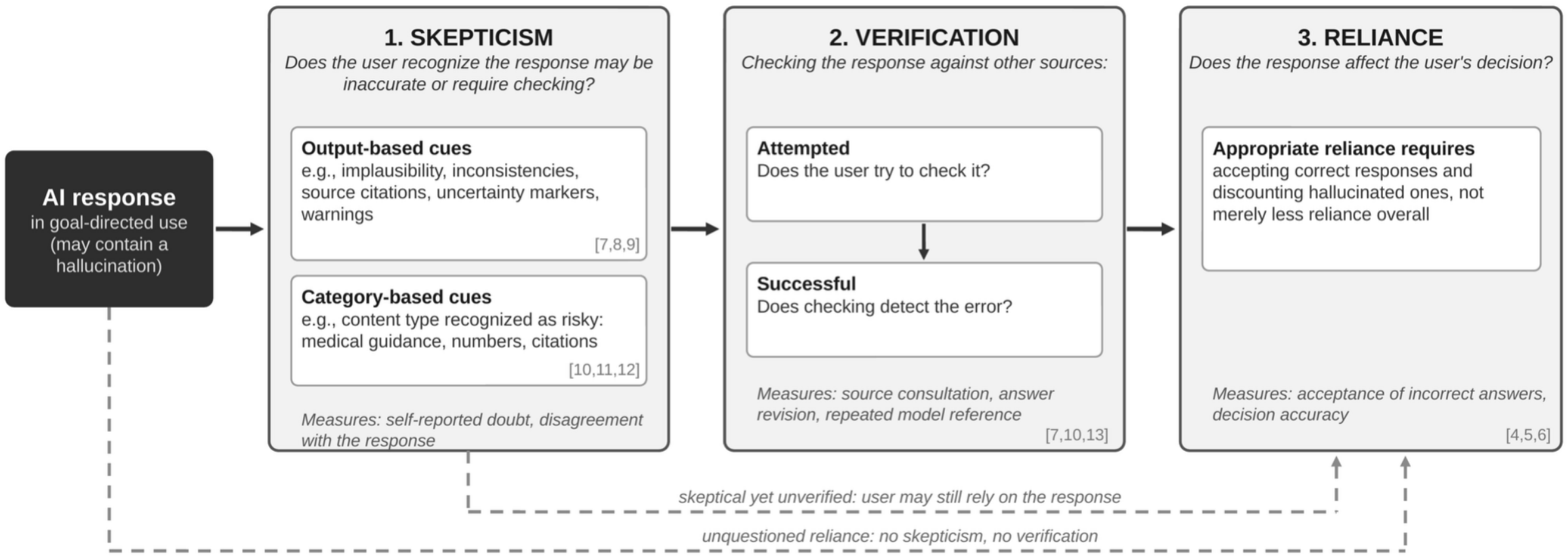

**Figure 1.** Constructs in the study of AI-advised decisions. Skepticism, verification, and reliance are conceptually distinct behavioral constructs: Users may be skeptical without verifying, verify without successfully detecting an error, or rely on a response without verifying it. Post-use reflections (satisfaction with the interaction, confidence in the answer) are retrospective appraisals. Table 1 classifies each reviewed study by the cue type manipulated (output-based vs. category-based) and by the constructs it examines (checkmark columns), using the definitions and examples shown here. Most studies also collect some measure of post-use reflection; because such measures are near-ubiquitous, we distinguish them conceptually here but do not track them as a separate focus in Table 1 or in the text.

Our review topic connects to research on misinformation, source evaluation, and correction, but the interactional setting differs from many standard paradigms in that literature. Users of organization-backed AI advisors are typically trying to complete a task rather than explicitly evaluate whether a claim is true. Spontaneous veracity assessments in cooperative communication are relatively infrequent absent an active trigger [14], and researchers have posited that consistent use of GenAI for task completion may lead users to delegate metacognitive oversight to the system, resulting in a state of "cognitive surrender" [15].

However, consumer tendencies to accept unverified outputs from organization-backed AI advisors, and the factors contributing to such acceptance, remain open questions.

Whether such generalizations extend to organization-backed AI advisors can be assessed by considering evidence that measures the antecedents of reliance, namely skepticism and verification, rather than reliance alone. As this review shows, data on organization-backed advisors are rare, and fewer studies have measured these antecedents directly. Establishing whether organization-backed AI advisors meaningfully reduce users' scrutiny relative to relevant baselines such as conventional search, non-AI information sources, or users' ordinary resources, therefore, requires evidence that the field has yet to produce. In the sections that follow, we review evidence on skepticism, verification, and reliance in goal-directed AI use; consider organizational and policy responses; and outline open methodological and psychological questions.

## 2. Skepticism, verification, and reliance in goal-directed use

Direct studies of consumers' skepticism toward, verification of, and reliance on hallucinated output from organization-backed AI advisors in real settings remain rare. Much of the available evidence, summarized in Table 1, comes from adjacent goal-directed paradigms, including AI-assisted search, question answering, simulated AI advice, content generation, and chatbot-assisted tasks. In these studies, AI-advised behavior is typically compared against baselines such as conventional search [17], users' ordinary resources [25], or a no-warning control [20]. Two features stand out about these studies. First, nearly every study includes some measure of reliance, whereas the antecedents of reliance – namely, skepticism and verification – are often left implicit and unmeasured. Second, the studies vary according to 1) the measured construct and 2) the type of cue manipulated to influence outcomes – that is, whether it originates in features of the response (output-based) or in the content category (category-based). We make several observations.

First, most studies that manipulate cues to prompt scrutiny of AI-generated information rely on output-based cues (i.e., cues present in the responses, such as inconsistencies or cited sources). Indeed, correct and hallucinated responses can share fluency, confidence, and rhetorical coherence, and thus output-based skepticism may depend most on content features that make a possible error visible, such as marking uncertain elements, quantifying confidence, or providing sources [29,30]. Among these, the more reliable effects arise from response features that lower verification costs. Source citations and noticeable inconsistencies are associated with reduced reliance on incorrect responses, whereas explanations alone can increase reliance on both correct and incorrect responses [7]. Sources lower the marginal cost of verification; explanations may instead make an answer feel complete without making it easier to evaluate. Other output-based cues that have been shown to move behavior are less easily translated to deployment, such as token-level highlighting (e.g., marker-style highlighting of parts of the response for which the model lacks confidence) or explicit statements that no source data exist, which an organization-backed AI advisor is unlikely to surface on its own. The output-based cue most feasible for organizations to implement, a general warning that errors are possible, tends to show weaker and more mixed evidence, a point the disclosure literature in Section 3 develops further.

Second, a smaller set of studies examine user scrutiny in specific content categories, asking whether users scrutinize an answer more when its subject matter is recognized as error-prone or high-stakes. Here, the evidence separates general from targeted information. Generic "ChatGPT can make mistakes" warnings did not alter diagnostic advice-taking relative to a no-warning control [20], and source-focused interventions increased general skepticism toward AI without reducing the influence of specific AI-generated misinformation [21]. More targeted information appears more promising: Descriptions of the model's behavior that identify where an AI system tends to fail improved error verification and appropriate reliance across three domains [10], and aligning large language model (LLM) explanations with the model's actual internal confidence narrowed the gap between user confidence and model accuracy [31].

Across these studies, one useful distinction is whether an intervention directs users' scrutiny toward specific higher-risk content rather than merely reminding them that errors are possible. A limitation runs through this group, however: Where content category is treated as the source of scrutiny, it is typically held fixed, with medical stimuli throughout, for example, rather than contrasted against a lower-risk category in the same design. Because the category is not varied, heightened scrutiny cannot be attributed to it, and the category-based pathway is assumed to operate rather than shown to do so.

Third, what the literature less often establishes is the process between cue and reliance. The manipulations are largely aimed at skepticism and verification, yet the outcomes recorded are most often reliance, so the intervening steps are often assumed from a downstream shift rather than observed directly. Few studies measure skepticism itself, whether in the moment or retrospectively, so whether skepticism moved is often inferred from a change in reliance. A related substitution occurs when self-report stands in for behavior. Participants in an LLM-search experiment reported a more satisfying experience with the LLM search tool even when the participants' decisions were less accurate [17]. AI advisor responses with explanations increased perceived response quality, reduced follow-up questions, and increased reliance on incorrect responses [7]. In organizational AI contexts, a branded interface may further complicate users' calibration, though this is understudied. The implication is that satisfaction with the interaction and self-reported confidence in the response are not sufficient proxies for appropriate reliance, and final decision accuracy is only an indirect proxy for skepticism and verification. Behavioral traces, including source consultation, answer revision, and repeated model reference, may offer a more direct window into these processes [13].

| | | | | | Focus | | |
|---|---|---|---|---|---|---|---|
| **Study** | **Setting and user goal** | **How AI error was introduced** | **Cue type** | **Behavioral outcome measured** | **Skepticism** | **Verification** | **Reliance** |
| **Buçinca et al. [16]** | Nutrition decisions with AI suggestions | Incorrect AI suggestions paired with cognitive forcing functions | Output-based (uncertainty, suggestion framing) | Disagreement with the AI; downstream decision accuracy | ✓ | | ✓ |
| **Spatharioti et al. [17]** | LLM-based vs. traditional search; consumer product choice (cars) | LLM occasionally provides incorrect product information | Output-based (highlight) | Final purchase-decision accuracy; search depth | ✓ | | ✓ |
| **Kim et al. [8]** | LLM-infused search engine; medical information seeking | Incorrect answers paired with first-person or impersonal uncertainty language | Output-based (uncertainty) | Self-reported reliance and answer accuracy | | ✓ | ✓ |
| **Kim et al. [7]** | LLM-infused application; objective question answering | Incorrect answers with or without explanations and sources | Output-based (explanations, sources) | Reliance on incorrect answers; follow-up questioning on confidence and verification | | ✓ | ✓ |
| **Cabrera et al. [10]** | AI-assisted tasks across three domains with behavior descriptions | AI errors with/without high-level failure-pattern descriptions | Category-based (Describing AI's accuracy for the context) | Error detection rate and appropriate reliance | | ✓ | ✓ |

| | | | | | Focus | | |
|---|---|---|---|---|---|---|---|
| **Study** | **Setting and user goal** | **How AI error was introduced** | **Cue type** | **Behavioral outcome measured** | **Skepticism** | **Verification** | **Reliance** |
| **Zhou et al. [18]** | Trivia answering with deployed LLMs | Confident incorrect model output, with and without epistemic markers | Output-based (confidence markers) | Choosing the model answer over independent lookup | | ✓ | ✓ |
| **Ashktorab et al. [19]** | Content-grounded data generation with an LLM assistant | Hallucinated source content embedded in the workflow | Output-based (formulate, read-first, highlight) | Whether users copy, append, edit, or correct AI text | | ✓ | ✓ |
| **Liu et al. [13]** | LLM-assisted tasks: quiz solving, article summarization, and trip planning | Plausible misinformation injected into LLM responses | Output-based (screen shows open search page) | Behavioral patterns: copy-paste, comprehension depth, repeated reference | | ✓ | ✓ |
| **Kıyak et al. [20]** | Medical students using ChatGPT for diagnostic reasoning | Standard "ChatGPT can make mistakes" warning vs. no warning | Output-based (general warning) | Diagnostic advice-taking behavior | | | ✓ |
| **Spearing et al. [21]** | Misinformation evaluation with AI-generated content | AI-generated misinformation with/without countermeasures | Output-based (sources; general disclaimer) | Belief accuracy and sharing intentions | | | ✓ |
| **He et al. [22]** | Knowledge-intensive tasks with AI assistance | AI outputs | Output-based (explanation) | Reliance on incorrect AI outputs conditional on perceived expertise | | | ✓ |
| **Salvi et al. [23]** | Debate interactions with GPT-4 on political topics | None | Category-based (personalized description) | Change in stated opinion pre- to post-interaction | | | ✓ |
| **Vasconcelos et al. [24]** | Maze-solving with simulated AI advice | Incorrect predictions accompanied by explanations | Output-based (explanations) | Acceptance of the incorrect AI prediction | | | ✓ |
| **Bo et al. [9]** | LLM advice on LSAT and numerical-estimation problems | Erroneous LLM outputs paired with reliance interventions | Output-based (disclaimer, highlighting, implicit answer) | Overreliance and appropriate reliance, with confidence calibration | | | ✓ |
| **Bean et al. [25]** | LLM as medical assistant for the general public; identify condition and disposition | Naturalistic LLM errors across ten clinical scenarios | None | Condition identification and disposition accuracy | | | ✓ |
| **Kaate et al. [26]** | AI-generated user persona for product research | Persona fabricates a response to an unanswerable question | Output-based (mentions of no data) | Whether the user adopts the fabricated response | | | ✓ |

| | | | | | Focus | | |
|---|---|---|---|---|---|---|---|
| **Study** | **Setting and user goal** | **How AI error was introduced** | **Cue type** | **Behavioral outcome measured** | **Skepticism** | **Verification** | **Reliance** |
| **Dell'Acqua et al. [27]** | Management consulting tasks (field experiment with Boston Consulting Group consultants) | Tasks inside vs. outside AI capability frontier | Category-based (prompt engineering training) | Quality of output; reliance on AI for tasks beyond its frontier | | | ✓ |
| **Pak et al. [28]** | Difficult quiz with chatbot assistant | Hallucinated answers interleaved with correct ones | Output-based (politeness) | Signal-detection sensitivity and response bias | | | ✓ |

**Table 1.** "Cue type" indicates whether the manipulated cue originates in features of the response (output-based) or in the content category (category-based). Checkmarks in the three "Focus" columns indicate which constructs each study examines; a study may examine more than one. Studies are grouped by primary focus: skepticism, then verification, then reliance. For Bo et al. [9], "implicit answer" denotes a presentation format in which the advice implies the answer without stating it directly, requiring users to extract the answer the advice supports.

## 3. Organizational and policy responses

We have examined how users are skeptical of, verify, and rely upon AI advisors in adjacent settings. We now ask how organizations handle hallucination risks. Organizations often mitigate hallucination risks through at least two kinds of responses, a system-side response and a user-side response, and the two act at different points.

The system-side response aims to reduce the frequency of hallucinations. For example, retrieval-augmented generation (RAG) architectures can, when properly designed, ground responses in verified knowledge bases, and human-in-the-loop review can provide a mechanism for independent verification before some outputs reach users [12,32]. Both can improve the information itself and operate before the user encounters it [33,34].

The user-side response does not primarily improve the underlying accuracy of the information output but shapes how the user responds to it. Disclaimers, interface labels, source citations, and uncertainty cues attempt to help users recognize when a response requires checking, to encourage verification, and to support appropriate reliance. Article 50 of the EU AI Act partly reflects this transparency logic, requiring that users be informed and that AI-generated content be identifiable [35].

We focus on user-side response for two reasons. First, its levers are the output-based and category-based cues we reviewed in Section 2, so asking whether they shift skepticism, verification, and reliance is the same question the reviewed evidence addresses. Second, reducing hallucinations does not remove the need to understand user-side responses. Even when models perform well on their own, assisted users often do not. For example, in the medical domain, lay users supported by frontier models identified relevant conditions and chose appropriate dispositions less often than the models scored in isolation, and no better than users relying on their ordinary resources, suggesting that the gap can arise from the interaction, not only from model-level error [25].

Among user-side responses, disclosure dominates, and it seems to mostly address skepticism. Disclaimers and labels aim to make users aware that any response may require checking, without specifying which responses or content categories warrant it. Yet the evidence that such warnings change behavior is weak and mixed: a general warning did not significantly increase advice-taking compared with a no-warning control [20], and a source-focused intervention increased general skepticism toward AI but did not significantly reduce the influence of specific AI-generated misinformation [21].

While disclosure may serve to limit the organization's liability, that protection appears limited. In *Moffatt v. Air Canada* [36], the tribunal rejected the airline's argument that its chatbot was a separate entity and held the organization responsible for misleading information delivered through its website, despite the availability of correct information elsewhere on the website. Courts in several jurisdictions have sanctioned attorneys for submitting AI-generated filings containing fabricated citations or other false authorities [37,38,39], framing those false authorities as failures of professional verification rather than as excusable tool errors [40]. Across these examples, responsibility appears to attach to how organizations and professionals verify AI-assisted outputs, not merely to whether the possibility of error was disclosed.

The broader record points the same way. Public databases of legal hallucination cases show that hallucinated materials continue to reach formal proceedings, suggesting that awareness, policies, or general warnings may be insufficient when verification remains optional, burdensome, or easy to bypass [41]. Similarly, AI hallucination-related risk disclosures in corporate governance materials indicate that AI error risk is increasingly being treated as an organizational issue rather than merely a user-interface design issue [42]. Organizations should be cautious about assuming that users will be skeptical or that verification will occur simply because users have been warned.

The same organization that posts a disclaimer also markets its system, and marketing can suppress the skepticism the disclaimer is meant to raise. Promoting an AI tool as reliable or already vetted invites users to treat its output as checked rather than as something to verify. In 2024, the U.S. Federal Trade Commission launched Operation AI Comply, targeting allegedly deceptive AI capability-related claims across legal, security, and content-generation tools [43,44,45], reflecting broader concerns about AI-washing and unsubstantiated capability claims. Whether an AI label raises or lowers skepticism is therefore an empirical question, and the answer is unlikely to be uniform.

The evidence reviewed here suggests that warnings and labels are unlikely, on their own, to make users skeptical of, or appropriately reliant on, the answers organization-backed AI advisors provide. The legal examples point in a similar direction: tribunals and courts have treated AI errors as matters of organizational or professional responsibility, not as problems solved by generic disclosure alone. Reducing reliance on hallucinations is therefore likely to depend less on warnings that errors are possible than on whether the interaction makes higher-risk content visible and verification feasible.

## 4. Open psychological, methodological, and organizational questions

Field studies of organization-backed AI advisors and how users respond hallucinations are rare, and the framework points to what such studies should measure. Reliance, the outcome these studies most often record, can follow from skepticism, attempted verification, or successful verification, and these should not be conflated. Work that aims to generalize to organization-backed AI advisors should measure skepticism, verification, and reliance separately. Because verification can be attempted without succeeding, these two layers should also be recorded separately. The reviewed evidence suggests, but does not test, that cue types may differ in which layer they affect: source citations appear to lower the cost of attempting verification [7], whereas descriptions of model behavior improved error detection, that is, verification success [10]. Whether output-based and category-based cues differentially influence verification success is itself an open question.

One noteworthy avenue for future research is that category-based skepticism (scrutinizing plausible output because the content type is recognized as higher risk) remains understudied. Users' ability and proclivity to scrutinize output from LLMs may depend on users' individual understanding of the competence boundaries of AI systems [11], although evidence from knowledge-work settings suggests that even domain experts may struggle to identify where AI systems are likely to perform poorly [27]. Future work should examine what forms of training, interface support, or task-specific feedback help improve recognition of higher-risk content categories over repeated interaction.

Additionally, empirical work on stable individual differences remains limited in this literature. Potentially relevant factors include prior experience, metacognitive self-monitoring, and confidence in both the AI system and one's task knowledge [16,46]. Existing findings suggest that confidence profiles may matter: Confidence in GenAI was associated with less critical thinking, while task-specific self-confidence was associated with more [46]. This suggests that skepticism and verification may depend not only on system design, but also on how users calibrate confidence in the tool and in their own task knowledge. Consumer research points to two relevant individual differences. Consumers with lower AI literacy are more receptive to AI because they are more likely to perceive it as magical [47]; the users least equipped to detect hallucinations may therefore approach AI advisors with the least scrutiny, though receptivity and reliance have yet to be studied together. Persuasion knowledge [48] likewise does not operate as a uniform form trait skepticism: accessing it can bolster credibility when an agent's tactic appears credible [2]. Whether users read the language of organization-backed AI advisors, such as source citations, disclaimers, and uncertainty expressions, as credible or manipulative tactics, and with what consequences for skepticism and verification, remains unstudied.

## References and recommended reading

Papers of particular interest, published within the period of review, have been highlighted as:

* of special interest

** of outstanding interest